\documentclass[aps,pre,showpacs,singlecolumn]{revtex4}
\usepackage{bm}
\usepackage{graphicx}
\bibstyle{approve.bib}
\usepackage{amssymb}
\usepackage{amsmath}
\usepackage{esint}
\usepackage{epstopdf}
\usepackage{color}
\usepackage{gensymb}
\usepackage{mathtools}
\usepackage{suffix}

\newcommand{\be}{\begin{equation}}

\newcommand{\ee}{\end{equation}}
\newcommand{\bea}{\begin{eqnarray}}
\newcommand{\eea}{\end{eqnarray}}
\newcommand{\lan}{\left\langle}
\newcommand{\ran}{\right\rangle}
\newcommand{\br}{\mathbf{r}}

\newcommand{\bk}{\mathbf{k}}

\newcommand{\e}{\varepsilon}

\newcommand{\tv}{\tilde{v}}

\newcommand{\pa}{\parallel}

\newcommand{\del}{\partial}
\newcommand{\lc}{\tilde{\lambda}_{\rm c}}
\newcommand{\md}{\mathrm{d}}
\newcommand{\f}{\bar{f}_0}

\begin{document}

\title{Dielectric trapping of biopolymers translocating through insulating membranes}

\author{Sahin Buyukdagli$^{1}$\footnote{email:~\texttt{buyukdagli@fen.bilkent.edu.tr}}, Jalal Sarabadani$^{2,4}$\footnote{email:~\texttt{jalal@ipm.ir}}, andTapio Ala-Nissila$^{3,4}$\footnote{email:~\texttt{Tapio.Ala-Nissila@aalto.fi}}}
\affiliation{$^{1}$Department of Physics, Bilkent University, Ankara 06800, Turkey\\
$^{2}$School of Nano Science, Institute for Research in Fundamental Sciences (IPM), 19395-5531, Tehran, Iran\\
$^{3}$Department of Applied Physics and QTF Center of Excellence, Aalto University School of Science, P.O. Box 11000, FI-00076 Aalto, Espoo, Finland\\
$^{4}$Interdisciplinary Centre for Mathematical Modelling and Department of Mathematical Sciences, Loughborough University, Loughborough, Leicestershire LE11 3TU, United Kingdom}

\begin{abstract}

Sensitive sequencing of biopolymers by nanopore-based translocation techniques requires extension of the time spent by the molecule in the pore. We develop an electrostatic theory of polymer translocation to show that the translocation time can be extended via the dielectric trapping of the polymer. In dilute salt conditions, the dielectric contrast between the low permittivity membrane and large permittivity solvent gives rise to attractive interactions between the cis and trans portions of the polymer. This self-attraction acts as a dielectric trap that can enhance the translocation time by orders of magnitude. We also find that electrostatic interactions result in the piecewise scaling of the translocation time $\tau$ with the polymer length $L$. In the short polymer regime $L\lesssim10$ nm where the external drift force dominates electrostatic polymer interactions, the translocation is characterized by the drift behavior $\tau\sim L^2$.  In the intermediate length regime $10\;{\rm nm}\lesssim L\lesssim\kappa_{\rm b}^{-1}$ where $\kappa_{\rm b}$ is the Debye-H\"{u}ckel screening parameter, the dielectric trap takes over the drift force. As a result, increasing polymer length leads to quasi-exponential growth of the translocation time. Finally, in the regime of long polymers $L\gtrsim\kappa_{\rm b}^{-1}$ where salt screening leads to the saturation of the dielectric trap, the translocation time grows linearly as $\tau\sim L$. This strong departure from the drift behavior highlights the essential role played by electrostatic interactions in  polymer translocation.

\end{abstract}


\date{\today}
\maketitle

\section{Introduction}

The continuous improvement of our control over nanoscale physics allows an increasingly broader range of nanotechnological applications for bioanalytical purposes.  Along these lines, the electrophoretic transport of biopolymers through nanopores can provide a surprisingly simple and fast approach for biopolymer sequencing~\cite{e1,e2,e3,e4,e5,e6,e7}. This sequencing technique consists in mapping the nucleic acid structure of the translocating polymer from the ionic current signal caused by the molecule. At present, the translocation times provided by experiments are not sufficiently long for sensitive reading of this ionic current signal~\cite{e6}. Thus, the technical challenge consists of reducing the polymer translocation speed by orders of magnitude from the current experimental values. Over the past two decade, this objective has motivated intensive research work with the aim to characterize the effect of various system characteristics on the polymer translocation dynamics.

Polymer translocation is driven by the entangled effects of electrostatic polymer-membrane interactions, the electrohydrodynamic forces associated with the electrophoretic and electroosmotic drags, and entropic barriers originating from conformational polymer fluctuations and hard-core polymer-membrane interactions. Due to the resulting complexity of the translocation process,  polymer translocation models have initially separately considered the contribution from electrohydrodynamic and entropic effects. Within Langevin dynamics, theoretical studies of polymer translocation first focused on the role played by entropy~\cite{n1,Luo,Sakaue1,n2,n5} (see also Refs.~\cite{Tapsarev,RevJalal} for an extended review of the literature). The contribution from electrostatics and hydrodynamics on the polymer translocation dynamics has been investigated by mean-field (MF) electrostatic theories~\cite{the3,the4,the6,the8}. Within a consistent electrohydrodynamic formulation, we have recently extended these translocation models by including beyond-MF charge correlations and direct electrostatic polymer-membrane interactions~\cite{Buy2015,Buy2017,Buy2018}.

In the theoretical modeling of polymer translocation, the current technical challenge consists of incorporating on an equal footing conformational polymer fluctuations and electrostatic effects. The achievement of this difficult task would allow to unify the entropic coarse-grained models and electrohydrodynamic theories mentioned above. At this point, it should be noted that such a unification necessitates the inclusion of polymer-membrane interactions outside the pore, while the translocation models of Refs.~\cite{Buy2015,Buy2017,Buy2018} developed for short polymers and long pores included exclusively the electrostatic polymer-membrane interactions inside the pore medium. In this work, we make the first attempt to overcome this limitation and develop a non-equilibrium theory of polymer translocation explicitly including the interactions between a charged dielectric membrane and an anionic polymer of arbitrary length. Within this theory, we characterize the effect of salt and membrane charge configurations, and the polymer length on the translocation dynamics of the molecule.

In Section~\ref{mod}, we introduce first the geometry and charge composition of the translocation model. Then, we derive the electrostatically augmented Fokker-Planck (FP) equation characterizing the translocation dynamics, and obtain the capture velocity and translocation time. Section~\ref{neut} considers the effect of surface polarization forces on polymer translocation through neutral membranes. Therein, we identify a \textit{dielectric trapping} mechanism enabling the extension of the translocation time by orders of magnitude. In Section.~\ref{char}, we investigate the effect of the fixed membrane charges on the dielectric trapping and reveal an \textit{electrostatic trapping} mechanism occuring at positively charged membranes in contact with physiological salt concentrations. We also scrutinize in detail the effect of dielectric and electrostatic interactions on the scaling of the polymer translocation time with the polymer length. Our results are summarized in the Discussion part where the limitations of our model and future extensions are discussed.

\section{Materials and Methods}
\label{mod}

\subsection{Charge composition of the system}
\label{chcom}

\begin{figure}
\hspace{2cm}\includegraphics[width=.68\linewidth]{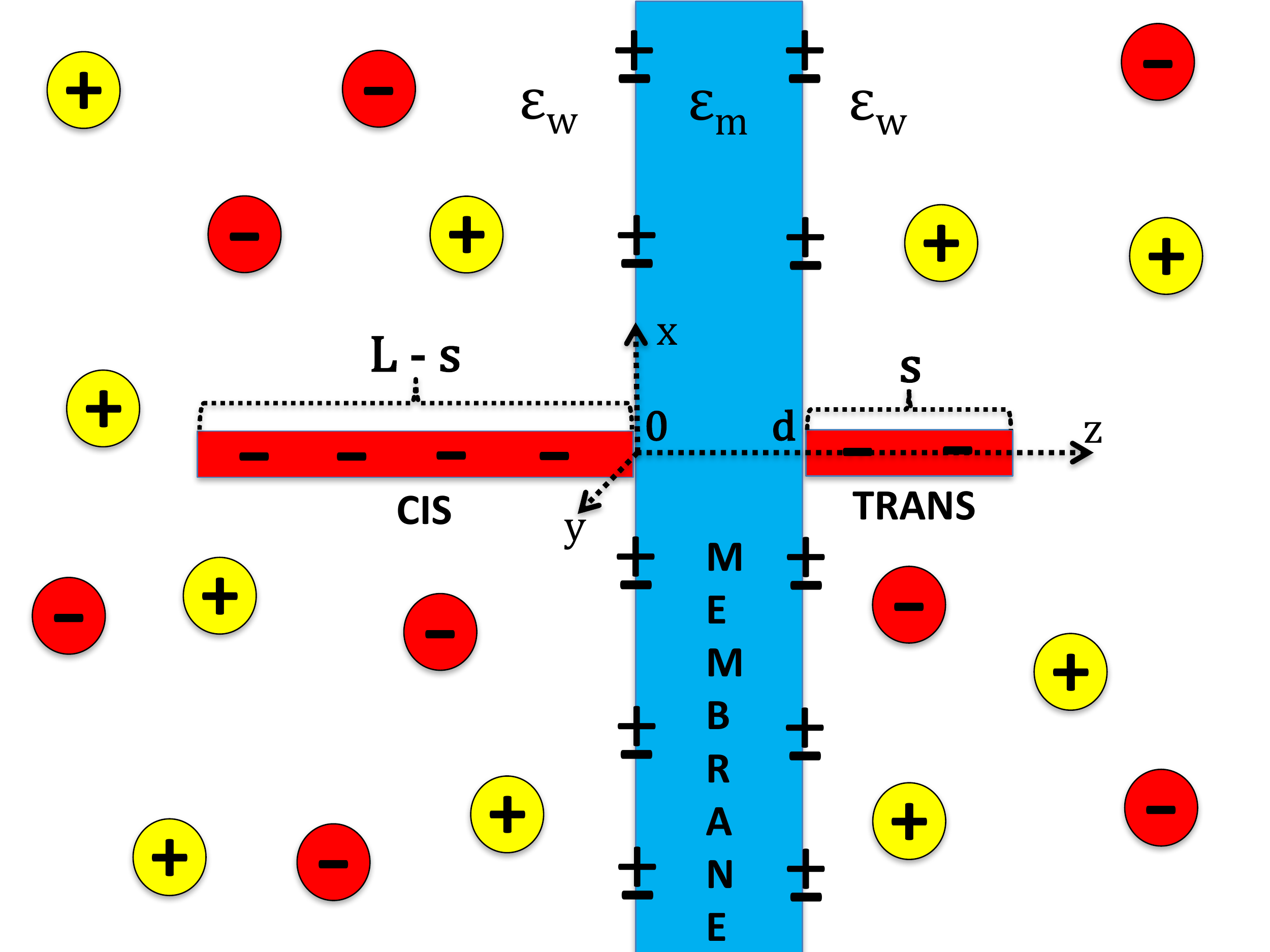}
\caption{(Color online) Schematic representation of the membrane of dielectric permittivity $\e_{\rm m}$, thickness $d$, and negative or positive surface charge density $\sigma_{\rm m}$. The membrane is immersed in the NaCl electrolyte with bulk density $\rho_{\rm b}$ and dielectric permittivity $\e_{\rm w}=80$. The polymer translocating through the pore has total length $L\gg d$. The length of the polymer portions on the \textit{cis} to the \textit{trans} sides is $L-s$ and $s$, respectively.}
\label{fig1}
\end{figure}

The charge composition of the system is depicted in Fig.~\ref{fig1}. The membrane of thickness $d$, surface charge $\sigma_{\rm m}$ of arbitrary sign, and dielectric permittivity $\e_{\rm m}$ is immersed in the monovalent electrolyte NaCl of concentration $\rho_{\rm b}$ and dielectric permittivity $\e_{\rm w}=80$. We note in passing that in our article, the dielectric permittivities are expressed in units of the vacuum permittivity $\e_0$. Moreover, the membrane contains a pore oriented along the $z$ axis. The externally applied voltage   between the \textit{cis} and \textit{trans} sides of the membrane induces a uniform electric field in the pore.  This field exerts a constant force $f_0$ on the polymer portion enclosed by the nanopore. 

The polymer is modeled as a charged line of total length $L$, mass $M$,  and the bare linear charge density of dsDNA molecules $-\lambda_{\rm c}q$ with $\lambda_{\rm c}=2.0/(3.4\;\mbox{{\AA}})$ and the electron charge $q=1.6\times10^{-19}$ C. Electrostatic polymer-membrane interactions induce an additional electrostatic force on the polymer charges. Appendix~\ref{ap3} explains the derivation of the corresponding electrostatic potential from the Debye-H\"{u}ckel (DH) level electrostatic polymer grand potential. The latter is obtained by expanding the grand potential of the whole system at the quadratic order in the polymer charge~\cite{Buy2016}. In order to improve this approximation, we will make use of the variational \textit{charge renormalization} technique~\cite{Netz} and evaluate the electrostatic polymer-membrane interactions in terms of the effective polymer charge density $\lc$ defined as
\be
\label{lc}
\lc=n\lambda_{\rm c}.
\ee
The effective charge density~(\ref{lc}) corresponds to the bare charge density $\lambda_{\rm c}$ dressed by the counterion cloud around the polyelectrolyte. In Eq.~(\ref{lc}), $n$ is the charge renormalization factor whose variational evaluation is explained in Appendix~\ref{varap}.  We finally note that in the limit of vanishing salt $\rho_{\rm b}\to0$, the charge renormalization factor tends to its \textit{Manning limit} $n=1/(\ell_{\rm B}\lambda_{\rm c})$~\cite{Netz} and the effective polymer charge~(\ref{lc}) becomes
\be
\label{man}
\lc=\frac{1}{\ell_{\rm B}},
\ee
where we used the Bjerrum length $\ell_{\rm B}=q^2/(4\pi\e_{\rm w}k_{\rm B}T)\approx7$ {\AA}, with the Boltzmann constant $k_{\rm B}$ and the ambient temperature $T=300$ K.

\subsection{Modified Fokker-Planck equation}
\label{MoF}

The reaction coordinate of the translocation is the length $s$ of the polymer portion on the \textit{trans} side. The polymer portion on the \textit{cis} side has length $L-s$ (see Fig.~\ref{fig1}). Thus, in our model, the contribution from the pore length to the translocation dynamics is neglected and  the right end of the polymer penetrating the membrane is assumed to reach immediately the \textit{trans} side. This is a reasonable approximation for the present case of thin membranes and long polymers $L\gg d$. This said, in the calculation of electrostatic polymer-membrane interactions, the finite thickness of the dielectric membrane will be fully taken into account.

The translocation dynamics is characterized by the Langevin equation
\be
\label{1}
\gamma M\frac{\md s}{\md t}=-\eta_{\rm p}\frac{\md s}{\md t}+f(s)+\xi(t),
\ee
where $\gamma$ is the hydrodynamic friction coefficient. The first term on the r.h.s. of Eq.~(\ref{1}) is the pore friction force and $\eta_{\rm p}$ the pore friction coefficient. The second term is the total external force $f(s)=-V'_{\rm p}(s)$ acting on the polymer, with the polymer potential $V_{\rm p}(s)$ including the effect of the externally applied electric force $f_0$ and electrostatic polymer-membrane interactions. Finally, the third term of Eq.~(\ref{1}) corresponds to the Brownian force $\xi(t)$. In the bulk electrolyte, the diffusion coefficient of a cylindrical molecule is given by~\cite{cyl1}
\be
\label{3}
D_{\rm b}=\frac{k_{\rm B}T}{3\pi\eta L}\ln\left(\frac{2L}{ea}\right),
\ee
with the water viscosity $\eta=8.91\times10^{-4}$ Pa s, Euler's number $e\approx2.718$, and the DNA radius $a=1$ nm. Thus, the corresponding hydrodynamic friction coefficient for the cylindrical molecule follows from Einstein's relation $MD_{\rm b}\gamma=k_{\rm B}T$ as 
\be
\label{4}
\gamma=\frac{3\pi\eta}{\lambda_{\rm m}\ln(2L/ea)},
\ee
where $\lambda_{\rm m}=M/L$ is the linear polymer mass density.

In Appendix~\ref{ap1}, we show that the effective FP equation associated with the Langevin Eq.~(\ref{1}) is given by
\be
\label{4II}
\del_t c(s,t)=D_{\rm p}\del_s^2 c(s,t)+\beta D_{\rm p}\del_s\left[U_{\rm p}'(s)c(s,t)\right],
\ee
where $c(s,t)$ is the polymer number density. In the dilute polymer regime where polymer-polymer interactions can be neglected, the function $c(s,t)$ also corresponds to the polymer probability density. In Eq.~(\ref{4II}), the effective pore diffusion coefficient is given by
\be\label{7}
D_{\rm p}=\frac{\gamma k_{\rm B}T}{M\gamma_{\rm p}^2},
\ee
with the net friction coefficient
\be\label{8}
\gamma_{\rm p}=\gamma+\frac{\eta_{\rm p}}{M}.
\ee
Finally, the effective polymer potential is
\be\label{9}
U_{\rm p}(s)=\frac{\gamma_{\rm p}}{\gamma}V_{\rm p}(s).
\ee

\subsection{Capture velocity $v_{\rm c}$ and translocation time $\tau$}
\label{vc}

We compute here the polymer translocation time $\tau$ and capture velocity $v_{\rm c}$. To this end, we express Eq.~(\ref{4II}) as an effective diffusion equation
\be
\label{5}
\frac{\del c(s,t)}{\del t}=-\frac{\del J(s,t)}{\del s},
\ee
with the polymer flux 
\be
\label{6}
J(s,t)=-D_{\rm p}\del_s c(s,t)-\beta D_{\rm p}U_{\rm p}'(s)c(s,t)
\ee
where the first and second terms on the r.h.s. correspond to the diffusive and convective flux components, respectively. We consider now the steady-state regime of the system characterized by a constant polymer flux $J(s,t)=J_{\rm st}$ and density $c(s,t)=c_{\rm st}(s)$. We recast Eq.~(\ref{6}) in the form
\be
\label{9II}
J_{\rm st}=-D_{\rm p}e^{-\beta U_{\rm p}(s)}\del_s\left[c_{\rm st}(s)e^{\beta U_{\rm p}(s)}\right].
\ee
Next, we integrate Eq.~(\ref{9II}) by imposing an absorbing boundary condition (BC) $c_{\rm st}(L)=0$ at the pore exit. The absorbing BC assumes that due to the deep voltage-induced electric potential on the trans side, the polymer that completes its translocation is removed from the system at $s=L$. One obtains
\be\label{10}
c_{\rm st}(s)e^{\beta U_{\rm p}(s)}=\frac{J_{\rm st}}{D_{\rm p}}\int_s^L\md s'e^{\beta U_{\rm p}(s')}.
\ee
Setting $s=0$ in Eq.~(\ref{10}), one gets the characteristic polymer capture velocity corresponding to the inward polymer flux per reservoir concentration $v_{\rm c}=J_{\rm st}/c_{\rm st}(0)$ as
\be
\label{11}
v_{\rm c}=\frac{D_{\rm p}}{\int_0^L\md s\;e^{\beta\left[U_{\rm p}(s)-U_{\rm p}(0)\right]}}.
\ee
We note that Eq.~(\ref{11}) corresponds to the characteristic speed at which the polymer reaches the minimum of the total electrostatic potential $V_p(z_p)$. In general, $v_c$ differs from the average translocation velocity. The capture and translocation velocities coincide only in the specific case of drift-driven translocation considered in Sec.~\ref{dri}.

In order to derive the translocation time, we first note that the polymer population in the pore follows from the integral of Eq.~(\ref{10}) in the form
\be
\label{12}
N_{\rm p}=\int_0^L\md s\;c_{\rm st}(s)=\frac{J_{\rm st}}{D_{\rm p}}\int_0^L\md s\;e^{-\beta U_{\rm p}(s)}\int_s^L\md s'\;e^{\beta U_{\rm p}(s')}.
\ee
The translocation time corresponds to the inverse translocation rate. The latter is defined as the polymer flux per total polymer number, i.e. $\tau^{-1}=J_{\rm st}/N_{\rm p}$. This gives the polymer translocation time in the form
\be
\label{13}
\tau=\frac{1}{D_{\rm p}}\int_0^L\md s e^{-\beta U_{\rm p}(s)}\int_s^L\md s' e^{\beta U_{\rm p}(s')}.
\ee
In Appendix~\ref{ap2}, we show that Eq.~(\ref{13}) can be also derived from the Laplace transform of the FP Eq.~(\ref{4II}) as the mean first passage time of the polymer from $s=0$ to $s=L$.

\subsection{Electrostatic polymer potential $V_{\rm p}(s)$}

The electrostatic potential experienced by the polymer reads
\be\label{2}
V_{\rm p}(s)=-f_0s+\Delta\Omega_{\rm p}(s).
\ee
The first term on the r.h.s. of Eq.~(\ref{2}) is the drift potential associated with the external force $f_0$. The second term including the polymer grand potential $\Delta\Omega_{\rm p}(s)$ accounts for electrostatic polymer-membrane interactions. In Appendix~\ref{ap3}, we show that this grand potential is given by
\be
\label{elpol}
\Delta\Omega_{\rm p}(s)=\Omega_{\rm pm}(s)+\Delta\Omega_{\rm intra}(s)+\Delta\Omega_{\rm inter}(s).
\ee
The first term on the r.h.s. of Eq.~(\ref{elpol}) corresponds to the direct interaction energy between the polymer and membrane charges,
\be\label{pmin}
\beta\Omega_{\rm pm}(s)=-\frac{2\lc}{\mu\kappa_{\rm b}^2}\left[2-e^{-\kappa_{\rm b}\left(L-s\right)}-e^{-\kappa_{\rm b}s}\right]\mathrm{sign}(\sigma_{\rm m}),
\ee
with the Gouy-Chapman length $\mu=1/(2\pi\ell_{\rm B}|\sigma_{\rm m}|)$ and DH screening parameter $\kappa_{\rm b}=\sqrt{8\pi\ell_{\rm B}\rho_{\rm b}}$. Then, the second term of Eq.~(\ref{elpol}) corresponding to the sum of the individual self interaction energies of the polymer portions on the \textit{cis} and \textit{trans} sides reads
\be
\label{cctt}
\beta \Delta\Omega_{\rm intra}(s)=\frac{\ell_{\rm B}\lc^2}{2}\int_0^\infty\frac{\mathrm{d}kk}{p_{\rm b}^3}\frac{\Delta\left(1-e^{-2kd}\right)}{1-\Delta^2e^{-2kd}}
\left\{\left[1-e^{-p_{\rm b}s}\right]^2+\left[1-e^{-p_{\rm b}(L-s)}\right]^2\right\},
\ee
where we defined the screening function $p=\sqrt{\kappa_{\rm b}^2+k^2}$ and the dielectric jump function $\Delta=(\e_{\rm w}p-\e_{\rm m}k)/(\e_{\rm w}p-\e_{\rm m}k)$. Finally, the interaction energy between the trans and \textit{cis} portions of the polymer is
\be
\label{ctin}
\beta \Delta\Omega_{\rm inter}(s)=\ell_{\rm B}\lc^2\int_0^\infty\frac{\mathrm{d}kk}{p_{\rm b}^3}\left\{\frac{\left(1-\Delta^2\right)e^{(p_{\rm b}-k)d}}{1-\Delta^2e^{-2kd}}-1\right\}e^{-p_{\rm b}d}\left[1-e^{-p_{\rm b}s}\right]\left[1-e^{-p_{\rm b}(L-s)}\right].
\ee

\section{Results and Discussions}

\subsection{Drift-driven regime}
\label{dri}

The drift-driven regime corresponds to the case of high salt density or strong external force $f_0$ where polymer membrane interactions can be neglected, i.e. $V_{\rm p}(s)\approx-f_0s$. In the drift limit, the effective polymer potential~(\ref{9}) takes the downhill linear form $\beta U_{\rm p}(s)=-\lambda_0s$ where we introduced the characteristic inverse length $\lambda_0=\beta f_0\gamma_{\rm p}/\gamma$. The capture velocity~(\ref{11}) and translocation time~(\ref{13}) become
\bea\label{14}
v_{\rm c}&=&\frac{D_{\rm p}\lambda_0}{1-e^{-\lambda_0L}},\\
\label{15}
\tau&=&\frac{1}{D_{\rm p}\lambda_0^2}\left[\lambda_0L-1+e^{-\lambda_0L}\right].
\eea
For strong electric forces with $\beta f_0L\gg1$, Eqs.~(\ref{14}) and~(\ref{15}) take the standard drift form
\bea\label{16}
v_{\rm c}&\approx&v_{\rm dr}=\frac{f_0}{\eta_{\rm p}+\lambda_{\rm m}\gamma L};\\
\label{17}
\tau&\approx&\tau_{\rm dr}=\frac{\eta_{\rm p}L+\lambda_{\rm m}\gamma L^2}{f_0},
\eea
satisfying the drift-driven transport equation $\tau\approx L/v_{\rm c}$. Considering that the logarithmic term in Eq.~(\ref{3}) is of order unity, and introducing the characteristic length $L_c=\eta_p/(3\pi\eta)$, Eq.~(\ref{17}) indicates that for short polymers $L\ll L_c$, the translocation time exhibits a linear dependence on the polymer length, i.e. $\tau\approx(\eta_{\rm p}/ f_0)L$. For long polymers $L\gg L_c$, the translocation time  grows quadratically with the polymer length as $\tau\approx(\lambda_{\rm m}\gamma/f_0)L^2$. We note that these scaling laws also follow from the rigid polymer limit of the tension propagation theory~\cite{n2}.

We verified that the translocation dynamics is qualitatively affected by the pore friction only in the drift-driven regime considered above. Thus, in order to simplify the analysis of the model, from now on, we switch off the pore friction and set $\eta_{\rm p}=0$. This yields in Eqs.~(\ref{8}) and~(\ref{9}) $\gamma_p=\gamma$.  Consequently, the effective polymer potential in Eqs.~(\ref{11}) and~(\ref{13}) becomes $U_p(s)=V_p(s)$ or
\be\label{17II}
U_{\rm p}(s)=-f_0s+\Omega_{\rm pm}(s)+\Delta\Omega_{\rm intra}(s)+\Delta\Omega_{\rm inter}(s).
\ee 

\subsection{Neutral membranes : dielectric trapping}
\label{neut}

We investigate here the electrostatics of polymer translocation through neutral membranes. In SiN membranes, the neutral surface condition is reached by setting the acidity of the solution to the isoelectric point value $\rm{pH}\approx5$~\cite{ph}. In this limit where $\sigma_{\rm m}=0$ and $\mu^{-1}=0$, the polymer-membrane coupling energy in the polymer potential~(\ref{17II}) vanishes, i.e. $\Omega_{\rm pm}(s)=0$. 
\begin{figure}
\includegraphics[width=.9\linewidth]{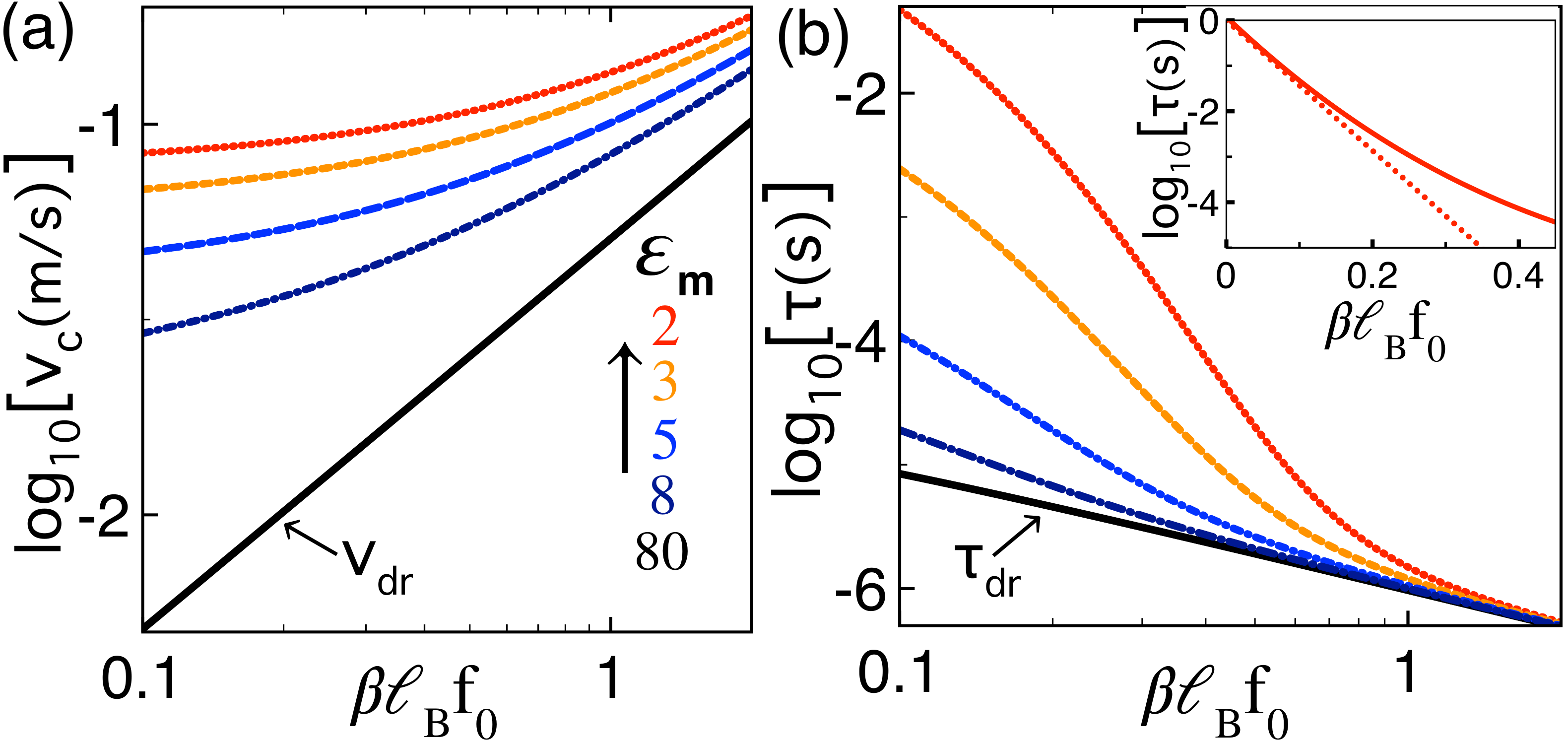}
\caption{(Color online) (\textbf{a}) Polymer capture velocity~(\ref{11}) and (\textbf{b}) translocation time~(\ref{13}) versus the dimensionless external force $\f=\beta\ell_{\rm B}f_0$ in the dilute salt regime $\kappa_{\rm b}=0$ and at various membrane permittivity values $\e_{\rm m}$ given in the legend of (a). The polymer length and membrane thickness are $L=50$ nm and $d=2$ nm. The pore friction is switched off, i.e. $\eta_{\rm p}=0$. The inset in (b) displays in a semilogarithmic plot the exponential regime of the translocation time for $\e_{\rm m}=2$.}
\label{fig2}
\end{figure}

\subsubsection{Dielectric trapping of the polymer in dilute salt}

To scrutinize the effect of polarization forces on the capture and translocation dynamics, we consider the simplest situation where the polymer is dressed by its counterions but there is no additional salt in the solvent, i.e. $\rho_{\rm b}=0$. This corresponds to the limit  $\kappa_{\rm b}\to0$ where the polymer self-energy components~(\ref{cctt}) and~(\ref{ctin}) become
\bea
\label{18}
\beta \Delta\Omega_{\rm intra}(s)&=&\frac{\Delta_0}{2\ell_{\rm B}}\int_0^\infty\frac{\mathrm{d}k}{k^2}\frac{1-e^{-2kd}}{1-\Delta_0^2e^{-2kd}}\left\{\left[1-e^{-ks}\right]^2+\left[1-e^{-k(L-s)}\right]^2\right\};\\
\label{19}
\beta \Delta\Omega_{\rm inter}(s)&=&-\frac{\Delta_0^2}{\ell_{\rm B}}\int_0^\infty\frac{\mathrm{d}k}{k^2}\frac{1-e^{-2kd}}{1-\Delta_0^2e^{-2kd}}e^{-kd}\left[1-e^{-ks}\right]\left[1-e^{-k(L-s)}\right],
\eea
with the dielectric parameter $\Delta_0=(\e_{\rm w}-\e_{\rm m})/(\e_{\rm w}+\e_{\rm m})$. According to Eqs.~(\ref{17II})-(\ref{19}), in the limit of vanishing dielectric discontinuity $\e_{\rm m}\to\e_{\rm w}$ where $\Delta_0=0$, polymer-membrane interactions disappear and one recovers the drift behavior of Eqs.~(\ref{16})-(\ref{17}).

In Figs.~\ref{fig2}(a) and (b), we display the polymer capture velocity $v_{\rm c}$ and translocation time $\tau$ against the dimensionless external force $\f=\beta\ell_{\rm B}f_0$ at various membrane permittivities $\e_{\rm m}\leq\e_{\rm w}$. One sees that in the weak external force regime $\f\lesssim1$, polarization effects arising from the low membrane permittivity result in the deviation of $v_{\rm c}$ and $\tau$ from the linear response behavior of Eqs.~(\ref{16}) and~(\ref{17}). More precisely, the external force dependence of the translocation time switches from linear $\tau\sim f_0^{-1}$ for large forces $\f\gtrsim1$ to exponential $\ln\tau\sim-f_0$ for weak forces $\f\lesssim0.2$ (see also the inset of Fig.~\ref{fig2}(b)). The exponential decay of $\tau$ with $f_0$ is the sign of the barrier-driven translocation that we scrutinize below. Figs.~\ref{fig2}(a) and (b) also show that at fixed force $f_0$, the dielectric discontinuity increases both the capture velocity $v_{\rm c}$ and the translocation time $\tau$ from their drift values, i.e. $\e_{\rm m}\downarrow v_{\rm c}\uparrow\tau\uparrow$.  The mutual enhancement of $v_{\rm c}$ and $\tau$ is an important observation for nanopore-based sequencing techniques whose efficiency depends on fast polymer capture and extended ionic current signal.

\begin{figure}
\includegraphics[width=.9\linewidth]{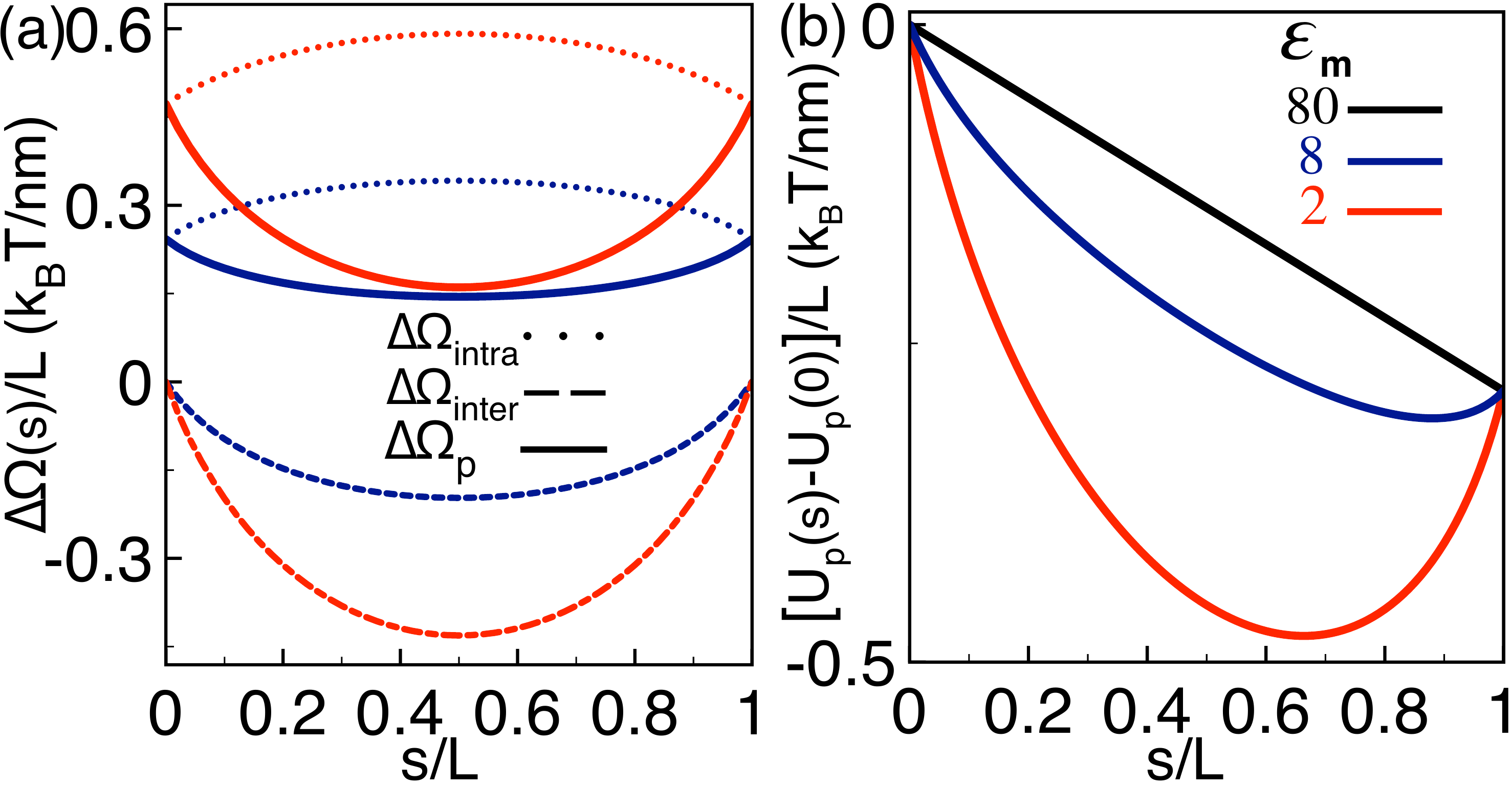}
\caption{(Color online) (\textbf{a}) Profiles of the polymer self-energy components Eq.~(\ref{18}) (dotted curves) and Eq.~(\ref{19}) (dashed curves), and the total self-energy in Eq.~(\ref{elpol}) (solid curves).  (\textbf{b}) Effective polymer potential Eq.~(\ref{17II}) renormalized by its value at $s=0$. In (a) and (b), the membrane permittivities are $\e_{\rm m}=80$ (black), $8$ (navy), and $2$ (red). The external force is $\f=0.2$. The other parameters are the same as in Fig.~\ref{fig2}.}
\label{fig3}
\end{figure}

The mechanism behind the enhanced capture speed and translocation time is illustrated in Figs.~\ref{fig3}(a) and (b). The plots display the electrostatic self-energy profiles, and the renormalized polymer potential $U_{\rm p}(s)-U_{\rm p}(0)$ that includes the electric force $f_0$ and determines the capture velocity~(\ref{11}) and translocation time~(\ref{13}).  First, we note that the self-energy component $\Delta\Omega_{\rm intra}(s)$ is concave and repulsive (dotted curves in Fig.~\ref{fig3}(a)). Thus, the individual image-charge interactions of the \textit{cis} and \textit{trans} portions of the polymer act as an electrostatic barrier limiting the polymer capture by the pore. Then, one sees that the energy component $\Delta\Omega_{\rm inter}(s)$ is convex and negative (dashed curves). Hence, the dielectric coupling between the \textit{cis} and \textit{trans} portions gives rise to an attractive force that favors the capture of the molecule.

In the present dilute salt conditions, the \textit{trans}-\textit{cis} coupling takes over the repulsive image-charge interactions. This gives rise to a purely convex and attractive total interaction potential $\Delta\Omega_{\rm p}(s)$ whose slope is enhanced with the magnitude of the dielectric discontinuity,  i.e. $\e_{\rm m}\downarrow|\Delta\Omega'_{\rm p}(s)|\uparrow$ (compare the solid curves in Fig.~\ref{fig3}(a)). Figure~\ref{fig3}(b) shows that as a result of this additional electrostatic force, the polymer potential develops an attractive well whose depth increases with the strength of the dielectric discontinuity, $\e_{\rm m}\downarrow\;\left[U_{\rm p}(s)-U_{\rm p}(0)\right]\downarrow$. This dielectrically induced potential well speeds up the polymer capture but also traps the polymer in its minimum, resulting in the mutual enhancement of the polymer capture speed and translocation time in Figs.~\ref{fig2}(a) and (b). 

In order to localize the position of the dielectric trap, we pass to the asymptotic insulator limit $\e_{\rm m}=0$ where the grand potential components~(\ref{18}) and~(\ref{19}) can be evaluated analytically as $\beta\Delta\Omega_{\rm intra}(s)=\ln(2)L/\ell_{\rm B}$ and
\be
\beta \Delta\Omega_{\rm inter}(s)=-\frac{L}{\ell_{\rm B}}\left\{\ln\left[\frac{L+d}{L+d-s}\right]+\frac{s}{L}\ln\left[\frac{L+d-s}{d+s}\right]+\frac{d}{L}\ln\left[\frac{d(d+L)}{(d+s)(d+L-s)}\right]\right\}.
\ee
Within this approximation, the solution of the equation $U'_{\rm p}(s_*)=0$ shows that the position of the trap rises linearly with the force $f_0$ and the polymer length $L$ as
\be
\label{tr}
s_*\approx\frac{1}{2}\left[L+\left(d+\frac{L}{2}\right)\f\right].
\ee
Equation~(\ref{tr}) can be useful to adjust the location of the dielectric trap in translocation experiments.

\begin{figure}
\includegraphics[width=.48\linewidth]{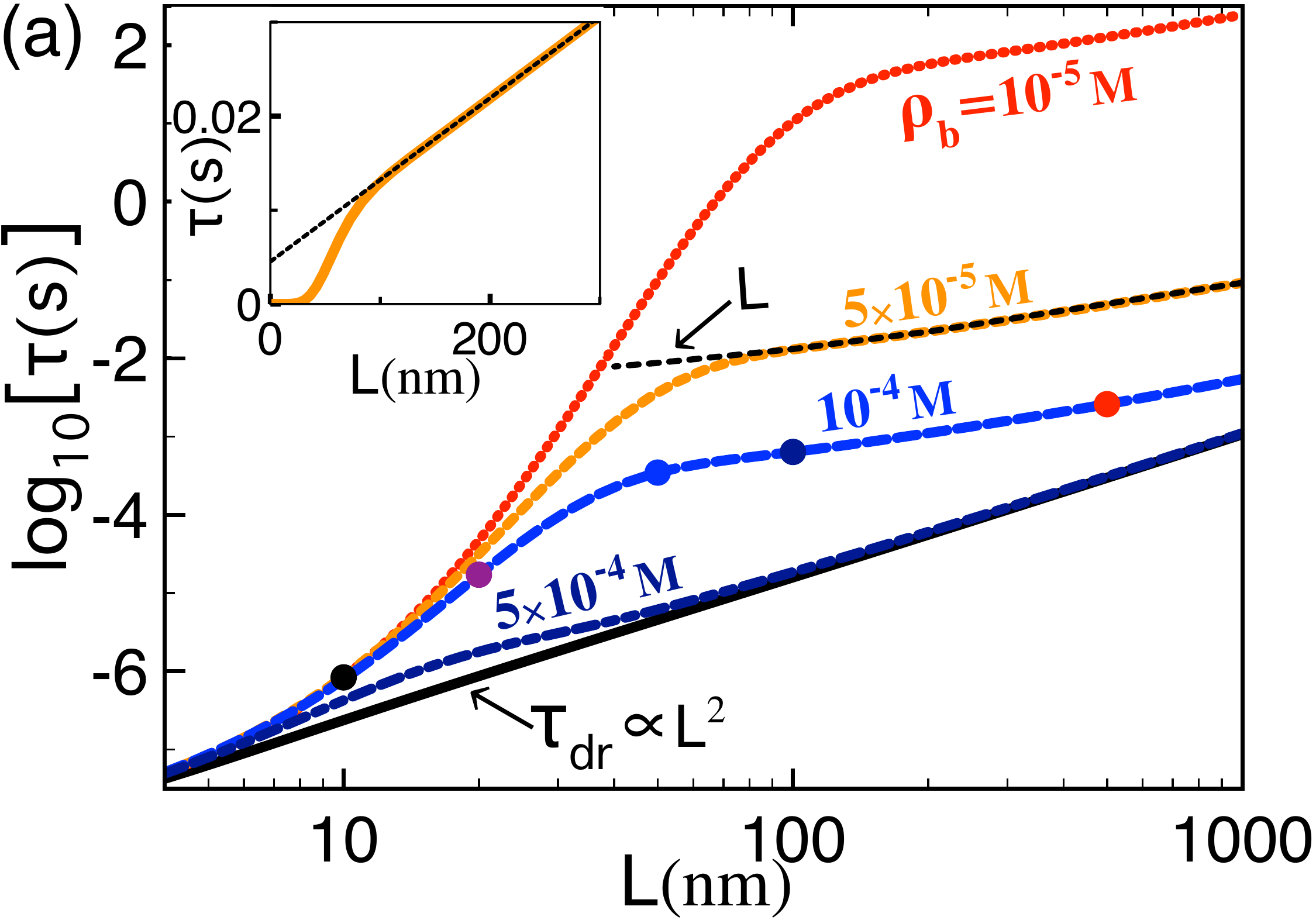}
\includegraphics[width=.5\linewidth]{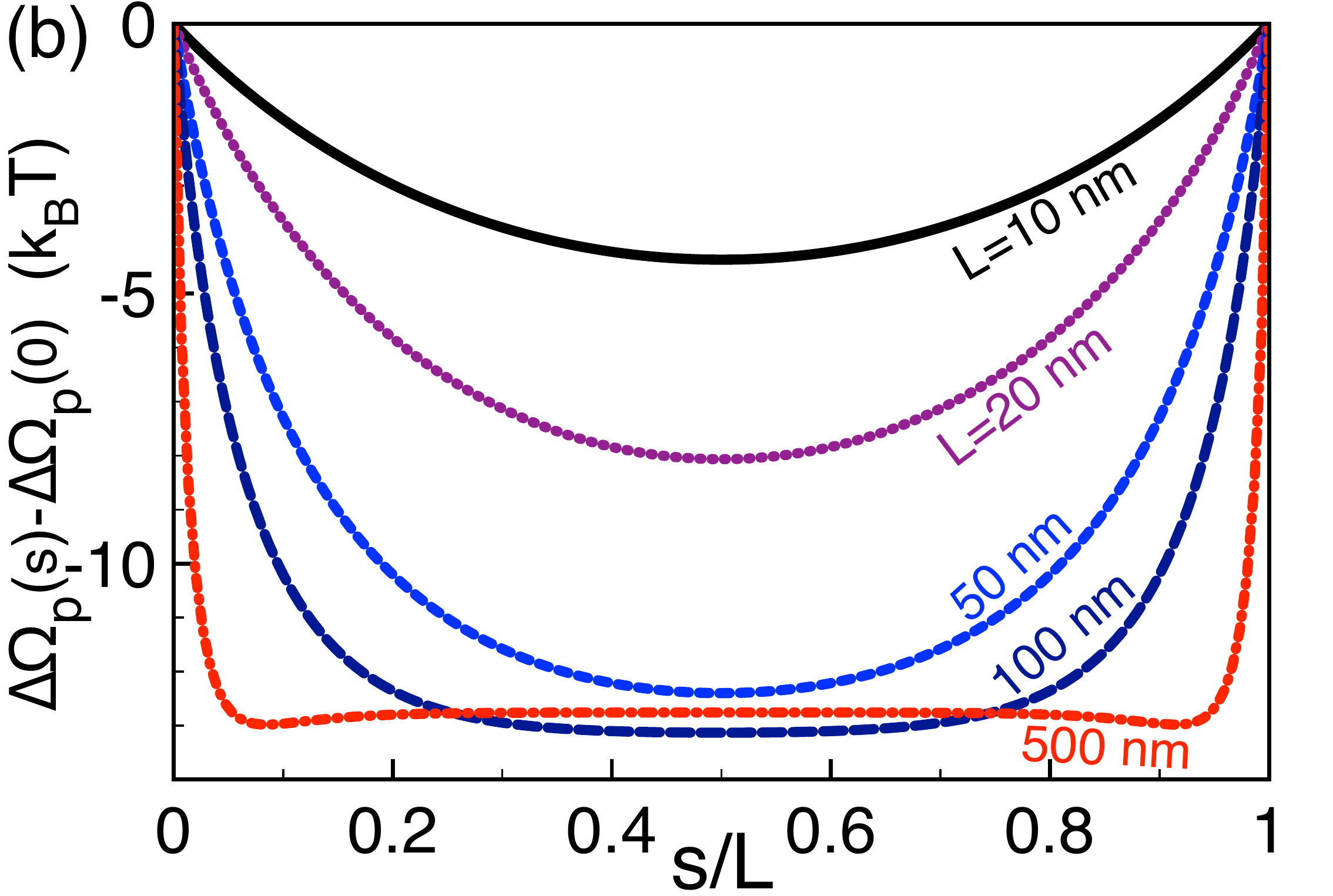}
\caption{(Color online) (\textbf{a}) Translocation time of Eq.~(\ref{13}) versus the polymer length at the membrane permittivity $\e_{\rm m}=2$ and various salt density values. The inset displays the translocation time at $\rho_{\rm b}=5\times10^{-5}$ M in a linear scale. (\textbf{b}) Renormalized polymer self-energy profile at the salt density $\rho_{\rm b}=10^{-4}$ M and various polymer lengths corresponding to the dots of the same color in (a). The external force and membrane thickness are $\f=0.2$ and $d=2$ nm.  In (a), the fast increase of the translocation time ends at the upper polymer length $L=L_+=\kappa_{\rm b}^{-1}$ with the numerical value $L_+=97$, $44$, $31$, and $20$ nm, for $\rho_{\rm b}=10^{-5}$, $5\times10^{-5}$, $10^{-4}$, and $5\times10^{-4}$ M.}
\label{fig4}
\end{figure}

\subsubsection{Effect of polymer length and finite salt concentration}

We scrutinize here the alteration of the polymer translocation time and capture speed by the polymer length and salt concentration. Figure~\ref{fig4}(a) shows that at a given salt concentration, the length dependence of the translocation time is characterized by three regimes. At short polymer lengths $L<L_-\approx10$ nm where polymer-membrane interactions and the self-energy $\Delta\Omega_{\rm p}(s)$ are weak, the translocation is characterized by drift transport, i.e. $\tau\approx\tau_{\rm dr}$. Consequently, the translocation time of short polymers rises quadratically with the molecular length, i.e. 
\be\label{sc1}
\tau\propto L^2\hspace{5mm}\mathrm{for}\hspace{2mm}L<L_-. 
\ee

The departure from drift transport occurs at intermediate lengths $L>L_-$. In this regime, the magnitude of the attractive \textit{trans}-\textit{cis} coupling becomes significant and the increase of the polymer length strongly enhances the depth of the electrostatic potential trap (see Fig.~\ref{fig4}(b)).  Figure~\ref{fig4}(a) shows that this results in the amplification of the translocation time with the polymer length by orders of magnitude. We found that this trend is the reminiscent of an exponential growth $\ln\tau\propto L$ reached in the asymptotic insulator limit $\e_{\rm m}=0$ (data not shown).

The quick rise of the translocation time with the polymer length continues up to the characteristic length $L\approx L_+=\kappa_{\rm b}^{-1}$ whose numerical value is given in the caption of Fig.~\ref{fig4}. Due to the salt screening of the \textit{trans}-\textit{cis} coupling, the depth of the dielectric trap is mostly invariant by the extension of the polymer length beyond $L_+$ (see Fig.~\ref{fig4}(b)). Thus, for $L\gtrsim L_+$,  the value of the double integral in Eq.~(\ref{13}) is not significantly affected by the length $L$, i.e. $\tau\propto D_{\rm p}^{-1}$. This results in the linear rise of the translocation time with the polymer length (see also the inset of Fig.~\ref{fig4}(a)), i.e. 
\be\label{sc2}
\tau\propto L\hspace{5mm}\mathrm{for}\hspace{2mm}L>L_+.
\ee

We note that the scaling discussed above qualitatively agrees with experiments on $\alpha$-Hemolysin pores exhibiting a similar piecewise length dependence of the translocation time (see e.g. Fig.9 of Ref.~\cite{Meller}). Finally, Fig.~\ref{fig4} shows that due to the screening  of dielectric polymer-membrane interactions, added salt reduces the translocation time, i.e. $\rho_{\rm b}\uparrow\tau\downarrow$. Beyond the characteristic salt concentration $\rho_{\rm b}\approx10^{-4}$ M where the length $L_+$ approaches $L_-$, the translocation time tends to its drift limit at all polymer lengths. 

\begin{figure}
\includegraphics[width=.9\linewidth]{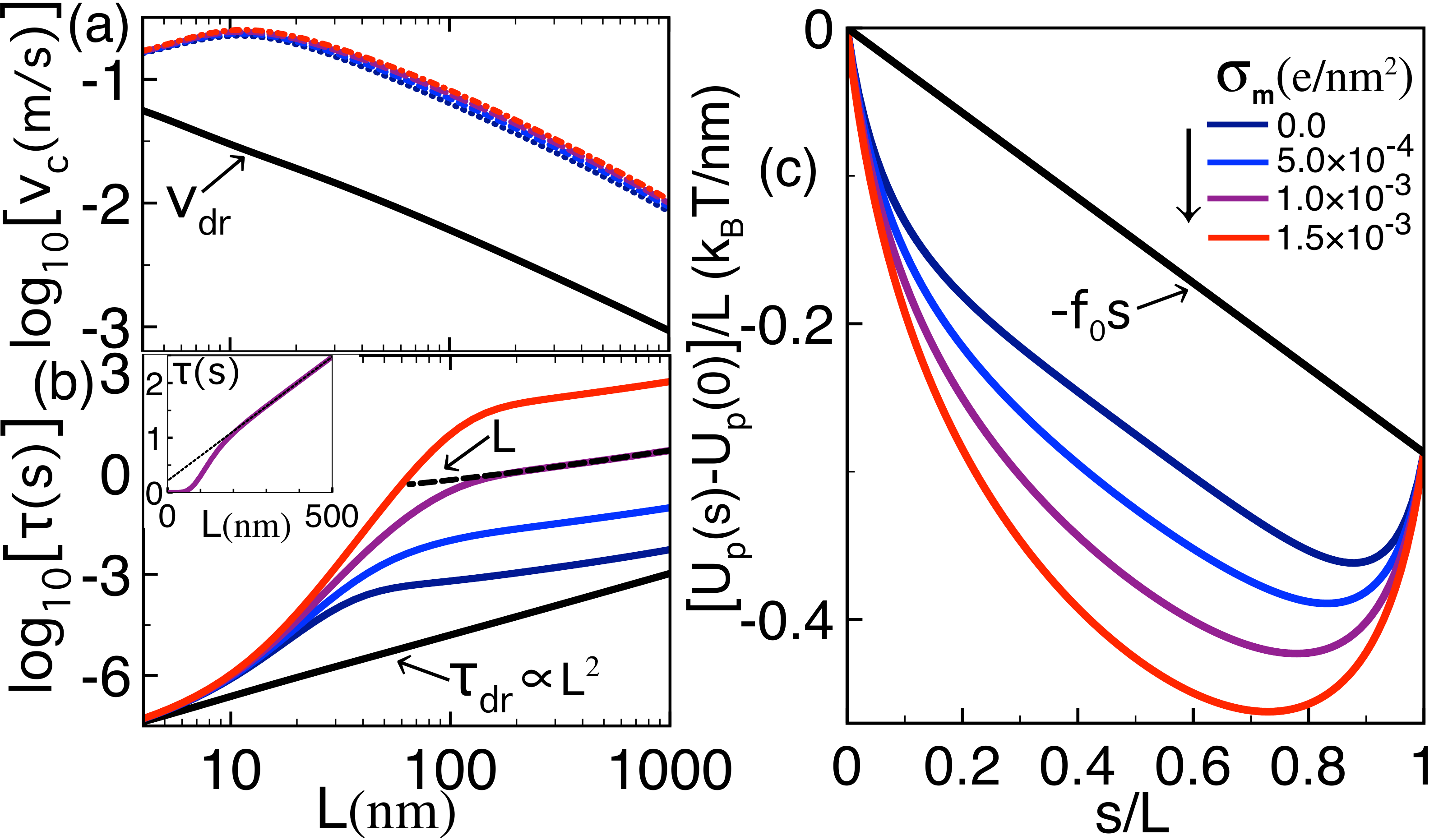}
\caption{(Color online) (\textbf{a}) Polymer capture velocity Eq.~(\ref{11}) and (\textbf{b}) translocation time Eq.~(\ref{13}) against the polymer length $L$. (\textbf{c}) Effective polymer potential Eq.~(\ref{17II}) at $L=100$ nm. The membrane charge density corresponding to each curve is given in the legend of (c). The inset in (b) displays the purple curve in a linear scale. Salt concentration is $\rho_{\rm b}=10^{-4}$ M. The other parameters are the same as in Fig.~\ref{fig4}.}
\label{fig5}
\end{figure}

\subsection{Charged membranes}
\label{char}

We investigate here the alteration of the features discussed in Sec.~\ref{neut} by a finite membrane charge.  For a positive membrane charge $\sigma_m\geq0$ corresponding to acidity values $\mbox{pH}\lesssim5$~\cite{ph}, the direct polymer-membrane coupling energy~(\ref{pmin}) results in an attractive force favoring the polymer capture. In order to characterize the effect of this additional force on the dielectric trapping mechanism, we first focus on the dilute salt regime and set $\rho_{\rm b}=10^{-4}$ M. Figures~\ref{fig5}(a)-(c) display the capture velocity, translocation time, and renormalized polymer potential at various weak membrane charge densities including the case of neutral membranes (navy curves). 

One first notes that upon the increase of the cationic membrane charge, the onset of the polymer-membrane attraction significantly deepens the trapping potential $U_p(s)-U_p(0)$. This enhances the translocation time of long polymers by orders of magnitude, i.e. $\sigma_{\rm m}\uparrow\tau\uparrow$ for $L\gtrsim30$ nm. However, one also sees that at the beginning of the translocation corresponding to the polymer capture regime $s\lesssim0.2\;L$, the slope of the polymer potential is weakly affected by the increment of the membrane charge density. As a result, the dielectrically enhanced capture velocity $v_c$ remains practically unaffected by a weak membrane charge.  Finally, Fig.~\ref{fig5}(b) shows that the linear scaling of the translocation time with the polymer length remains unchanged by  the surface charge, i.e. $\tau\propto L$ for $L\gtrsim L_+$ (see also the inset). One however notes that the finite membrane charge shifts the regime of linearly rising translocation time to larger polymer lengths, i.e. $\sigma_m\uparrow L_+\uparrow$.

\begin{figure}
\includegraphics[width=.9\linewidth]{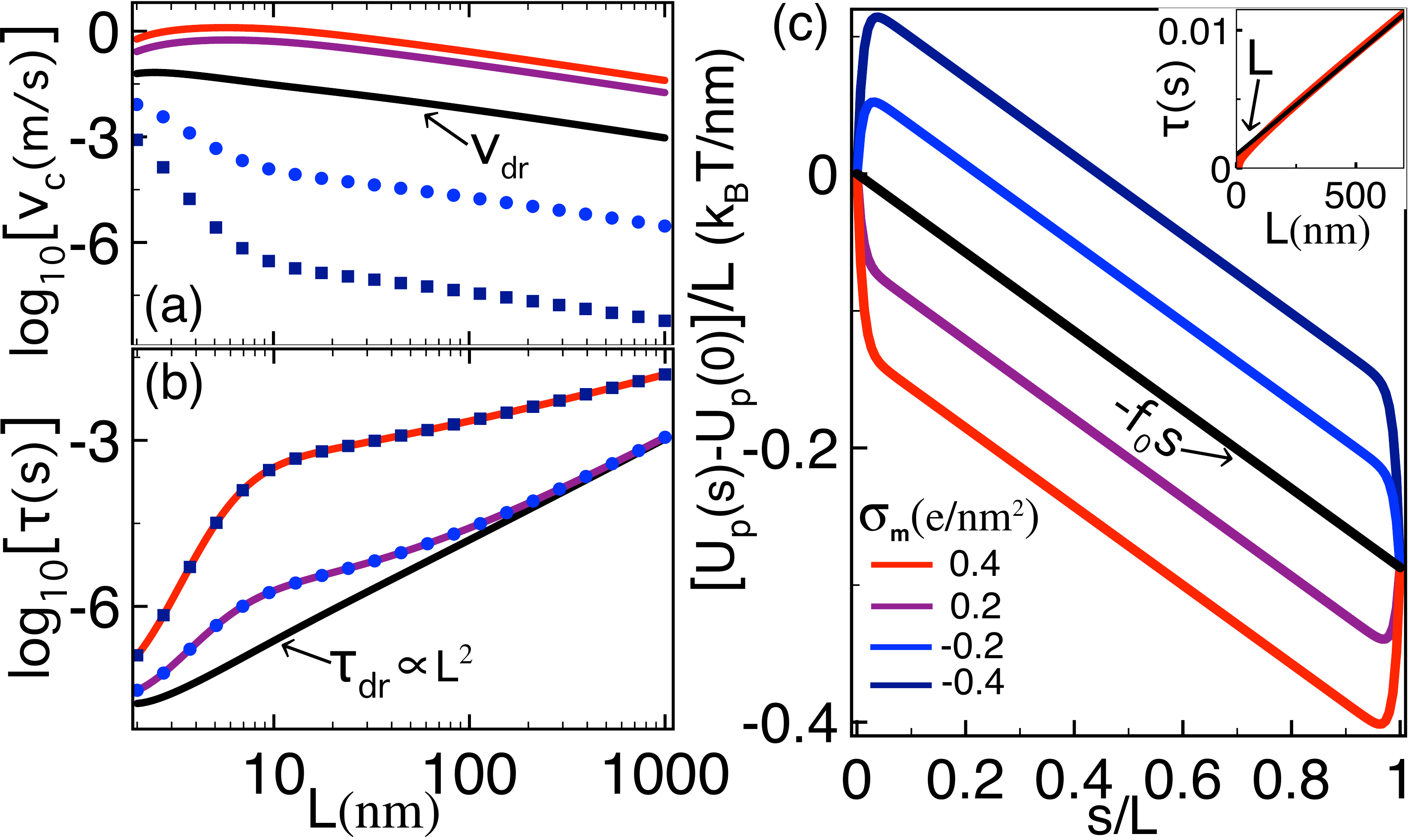}
\caption{(Color online) (\textbf{a}) Polymer capture velocity Eq.~(\ref{11}) and (\textbf{b}) translocation time Eq.~(\ref{13}) against the polymer length $L$. (\textbf{c}) Effective polymer potential~(\ref{17II}) at $L=100$ nm. The inset in (c) displays the red curve of (b) in a linear scale. The membrane charge density corresponding to each curve is given in the legend of (c). Salt concentration is $\rho_{\rm b}=0.1$ M. The other parameters are the same as in Fig.~\ref{fig4}.}
\label{fig6}
\end{figure}

We consider now the stronger salt regime where the dielectric trapping effect disappears. To simplify the numerical computation of the capture velocity Eq.~(\ref{11}) and translocation time Eq.~(\ref{13}), we neglect the dielectric interaction terms of Eq.~(\ref{17II}) that become perturbatively small. Within this approximation, the polymer potential becomes $U_{\rm p}(s)\approx-f_0s+\Omega_{\rm pm}(s)$. 

Figures~\ref{fig6}(a)-(c) show that in the regime of moderate salt concentration and cationic membrane charge (purple and red curves), the direct polymer-membrane charge attraction can solely induce a deep enough electrostatic trap to enhance both the capture velocity and the translocation time by orders of magnitude, i.e. $\sigma_m\uparrow v_c\uparrow\tau\uparrow$. In terms of the dimensionless constant $c=\beta\mu\kappa_{\rm b}f_0/(4\lc)$, the relation $U'_{\rm p}(s_*)=0$ yields the location of the electrostatic trap in the form
\be\label{tr2}
\frac{s^*}{L}=1+\frac{1}{\kappa_{\rm b}L}\ln\left(c+\sqrt{c^2+e^{-\kappa_{\rm b}L}}\right).
\ee
Equation~(\ref{tr2}) can enable to control the position of the polymer trap by changing the relative weight of the drift force $f_0$ and the electrostatic polymer-membrane attraction via the adjustment of various system parameters. Indeed, in the regime $\beta\mu\kappa_{\rm b}f_0/(4\lc)\ll1$ corresponding to weak salt or external force $f_0$, and high membrane charge $\sigma_{\rm m}$ and/or polymer charge $\lc$,  Eq.~(\ref{tr2}) indicates the trapping of the polymer at $s^*\approx L/2$. Moving to the opposite regime $\beta\mu\kappa_{\rm b}f_0/(4\lc)\gg1$ of strong salt or external force, low membrane or polymer charge strength, and long polymers $\kappa_{\rm b}L\gtrsim1$, the trapping point in Eq.~(\ref{tr2}) is shifted towards the polymer exit $s^*=L$ according to the relation
\be\label{tr3}
\frac{s^*}{L}\approx1-\frac{1}{\kappa_{\rm b}L}\ln\left(\frac{2\lc}{\beta\mu\kappa_{\rm b}f_0}\right).
\ee
Figure~\ref{fig6}(b) also shows that at strong enough membrane charges (red curve), the trapping-induced enhancement of the translocation time is followed at large lengths by the linear scaling behavior $\tau\propto L$ equally observed for neutral and weakly charged membranes (see the inset of Fig.~\ref{fig6}(c)). At intermediate charges, the system tends to the drift behavior $\tau\to\tau_{\rm dr}$ before the linear scaling regime is reached (purple curve in Fig.~\ref{fig6}(b)).

We finally investigate the effect of anionic membrane charges reached in the acidity regime $\mbox{pH}\gtrsim5$. Interestingly, Fig.~\ref{fig6}(b) indicates that in strong salt conditions where the polymer-membrane charge coupling dominates the dielectrically induced polymer self-interactions, the enhancement of the translocation time in Eq.~(\ref{13}) does not depend on the sign of the membrane charge, i.e. $\tau(\sigma_m)=\tau(-\sigma_m)$. However, one also notes that in anionic membranes, the like-charge polymer-membrane repulsion gives rise to an electrostatic barrier at the pore entrance (see Fig.~\ref{fig6}(c)). Fig.~\ref{fig6}(a) shows that this barrier diminishes the polymer capture rate by several orders of magnitude, i.e. $|\sigma_m|\uparrow U_p(s)-U_p(0)\uparrow v_c\downarrow$ for $\sigma_m<0$. Thus, in anionic membranes, the enhancement of the translocation time stems from the suppression of polymer capture by the electrostatic polymer-membrane repulsion. The existence of a similar barrier induced by electrostatic DNA-pore repulsion has been previously identified by a different polymer translocation model developed for long nanopores and short polymers~\cite{Buy2017}.

\section{Discussion}

The accurate characterization of voltage-driven polymer translocation requires modeling of this process by including the electrostatic details of the polymer-membrane complex and the surrounding electrolyte solution. Motivated by this need, we have developed here an electrostatic transport model to investigate the effect of surface polarization forces, added salt, and membrane charge on the capture and translocation of stiff polymers with arbitrary length. Our results are summarized below.

We first considered the case of neutral membranes and dilute salt regime where the polyelectrolyte is dressed by its counterions but there is no additional salt in the system. In this regime, we identified a dielectrically induced polymer trapping mechanism. Namely, the dielectric contrast between the low permittivity membrane and large permittivity solvent leads to attractive interactions between the cis and trans portions of the polymer.  The attraction gives rise to a \textit{dielectric trap} located at $s=s^*=[L+(d+L/2)\f]/2$. The trap speeds up the polymer capture occurring at $s<s^*$ but slows down the escape of the polymer at $s>s^*$, amplifying the polymer capture velocity by several factors and the total translocation time by orders of magnitude.

We also observed that in neutral membranes, added salt of concentration $\rho_{\rm b}\gtrsim10^{-4}$ M suppresses the dielectric trapping of the polymer. However, at arbitrary salt densities, positive membrane surface charges emerging at low solution pH restore the polymer trapping via the electrostatic polymer-membrane attraction. This \textit{electrostatic trap} can enhance the polymer capture speed and translocation time as efficiently as its dielectric counterpart. It was also shown that the location of the trap in Eq.~(\ref{tr2}) can be adjusted by modifying the experimentally accessible model parameters such as the salt and membrane charge density. Thus, the electrostatic trapping can equally well provide an efficient way to extend the duration of the ionic current blockade required for the sensitive sequencing of the translocating biopolymer.

Finally, we investigated the effect of polymer trapping on the scaling of the translocation time with the polymer length. At short lengths $L\lesssim10$ nm where the interactions between the cis and trans sides of the polymer are dominated by the drift force $f_0$, the translocation is characterized by the drift behavior of Eq.~(\ref{17}). In the intermediate polymer length regime $10\;{\rm nm}\lesssim L\lesssim\kappa_{\rm b}^{-1}$ where the attractive trans-cis coupling takes over the drift force, the resulting dielectric trap leads to a quasi-exponential inflation of the translocation time with the length of the molecule.  Beyond the characteristic polymer length $L\approx\kappa^{-1}_{\rm b}$ where ionic screening comes into play, the depth of the dielectric trap saturates. As a result, the translocation time of long polymers rises linearly with the molecular length, i.e. $\tau\propto L$. We finally showed that in positively charged membranes, the electrostatic trap results in a similar piecewise length dependence of the translocation time. It is also important to note that such a piecewise trend has been previously observed in translocation experiments with $\alpha-$Hemolysin pores~\cite{Meller}.

The present formalism developed for long polymers and thin membranes is complementary to our previous translocation model of Ref.~\cite{Buy2017} introduced for short polymers translocating through long nanopores. These two formalisms can be unified in the future by taking into account both the detailed electrohydrodynamics of the nanopore and electrostatic polymer-membrane interactions outside the pore. This extension would also enable to consider the influence of non-linear electrostatic correlation effects such as  polymer and pore charge inversion on the translocation dynamics~\cite{Buy2015}. Finally, the inclusion of entropic polymer fluctuations will allow to incorporate into our electrostatic formalism the tension propagation mechanism relevant for long polymers~\cite{Sakaue1,RevJalal}.

\appendix

\section{Variational evaluation of the dressed polymer charge}
\label{varap}

We summarize here the variational \textit{charge renormalization} procedure that allows to evaluate the effective polymer charge density $\lc$. The latter is defined as $\lc=n\lambda_{\rm c}$, where the charge renormalization factor $n$ accounting for the counterion dressing of the bare charge $\lambda_{\rm c}$ follows from the numerical solution of the variational equation~\cite{Netz}
\be
\label{var}
2(1-n)\ell_{\rm B}\lambda_{\rm c}\psi_{\rm p}(a)+\kappa_{\rm b}^2\int_a^\infty\mathrm{d}rr\left\{n\psi_{\rm p}^2(r)-\psi_{\rm p}(r)\sinh\left[n\psi_{\rm p}(r)\right]\right\}=0.
\ee
In Eq.~(\ref{var}), the radius  of the cylindrical polymer is $a=1$ nm, and the electrostatic potential induced by the bare polymer charge reads
\be
\label{psm}
\psi_{\rm p}(r)=\frac{2\ell_{\rm B}\lambda_{\rm c}}{\kappa_{\rm b}a}\frac{\mathrm{K}_0(\kappa_{\rm b}r)}{\mathrm{K}_1(\kappa_{\rm b}a)}.
\ee
In Eq.~(\ref{psm}), we used the DH screening parameter $\kappa_{\rm b}=\sqrt{8\pi\ell_{\rm B}\rho_{\rm b}}$ and the modified Bessel functions $\mathrm{K}_n(x)$~\cite{math}. We also note that in the \textit{Manning limit} of vanishing salt $\kappa_{\rm b}\to0$ where Eq.~(\ref{var}) yields $n=1/(\ell_{\rm B}\lambda_{\rm c})$, the net polymer charge becomes $\lc=1/\ell_{\rm B}$.

\section{Derivation of the FP equation~(\ref{4II})}
\label{ap1}

In this appendix, we derive the FP Eq.~(\ref{4II}) associated with the Langevin Eq.~(\ref{1}). First, we cast this equation in the form
\be
\label{I}
\gamma_{\rm p} M\frac{\md s}{\md t}=f(s)+\xi(t),
\ee
with the net friction coefficient
\be\label{II}
\gamma_{\rm p}=\gamma+\frac{\eta_{\rm p}}{M},
\ee
and the Gaussian white noise satisfying the relations
\bea\label{III}
&&\lan\xi(t)\ran=0;\\
\label{IV}
&&\lan\xi(t)\xi(t')\ran=2M\gamma k_{\rm B}T\delta(t-t').
\eea
In Eqs.~(\ref{III})-(\ref{IV}), the bracket $\lan\cdot\ran$ indicates the average over the Brownian noise. Integrating Eq.~(\ref{I}) over the infinitesimal time interval $\delta t$, one gets
\be
\label{V}
\delta s(t)=\frac{1}{M\gamma_{\rm p}}f(s)\delta t+\frac{1}{M\gamma_{\rm p}}\int_t^{t+\delta t}\md t'\xi(t').
\ee
Taking the noise average of Eq.~(\ref{V}) and its square, and keeping only the terms linear in $\delta t$, one obtains
\bea
\label{VI}
\lan\delta s(t)\ran&=&\frac{1}{M\gamma_{\rm p}}f(s)\delta t;\\
\label{VII}
\lan\delta s^2(t)\ran&=&\frac{2\gamma k_{\rm B}T\delta t}{M\gamma_{\rm p}^2}.
\eea

We now derive the stochastic equation generating the averages~(\ref{VI}) and~(\ref{VII}). Following the approach of Ref.~\cite{FP}, we start with the Chapman-Kolmogorov equation for the polymer translocation between the initial position $s_0=s(t_0)$ and final position $s=s(t+\delta t)$,
\be
\label{VIII}
c(s,t+\delta t;s_0,t_0)=\int_0^L\md s'c(s,t+\delta t;s',t)c(s',t;s_0,t_0).
\ee
To progress further, we express the noise-averaged definition of the probability density 
\be\label{X}
c(s,t+\delta t;s',t)=\lan\delta(s-s'-\delta s)\ran,
\ee
where the term $\delta s$ on the r.h.s. corresponds to the random displacement over the infinitesimal time interval $\delta t$. Next, we Taylor-expand the corresponding term at order $O\left(\delta s^2\right)$ to obtain
\be\label{XI}
c(s,t+\delta t;s',t)=\left\{1+\lan\delta s\ran\del_{s'}+\frac{1}{2}\lan\delta s^2\ran\del^2_{s'}\right\}\delta(s-s').
\ee
Inserting Eq.~(\ref{XI}) into Eq.~(\ref{VIII}), carrying out integrations by parts, and using the relations~(\ref{VI}) and~(\ref{VII}), one obtains
\be\label{XI2}
c(s,t+\delta t;s_0,t_0)=\left\{1-\frac{\delta t}{M\gamma_{\rm p}}\del_{s}f(s)+\delta tD_{\rm p}\partial_s^2\right\}c(s,t;s_0,t_0),
\ee
where we introduced the effective diffusion coefficient 
\be\label{XII}
D_{\rm p}=\frac{\gamma k_{\rm B}T}{M\gamma_{\rm p}^2}.
\ee
Taylor-expanding the l.h.s. of Eq.~(\ref{XI}), one gets
\be\label{XIII}
c(s,t+\delta t;s_0,t_0)=\left\{1+\delta t\del_t\right\}c(s,t;s_0,t_0).
\ee
Equating the relations~(\ref{XI}) and~(\ref{XIII}) and simplifying the result, one finally obtains the modified FP equation
\be\label{fp1}
\del_t c(s,t)=D_{\rm p}\del_s^2 c(s,t)+\beta D_{\rm p}\del_s\left[U_{\rm p}'(s)c(s,t)\right]
\ee
including the effective polymer potential 
\be\label{XIV}
U_{\rm p}(s)=\frac{\gamma_{\rm p}}{\gamma}V_{\rm p}(s).
\ee

\section{Calculation of the translocation time}
\label{ap2}

Here, based on the FP Eq.~(\ref{fp1}), we derive the polymer translocation time~(\ref{12}) as the mean first passage time of the polymer from the \textit{cis} to the \textit{trans} side. Our derivation will follow the approach of Ref.~\cite{Ansalone} that will be extended to the presence of a steady-state solution to the FP Eq. The BCs associated with this equation are the initial condition at the pore mouth and an absorbing boundary at the pore exit,
\bea
\label{a2}
c(s,t=0)&=&\delta(s);\\
\label{a3}
c(s=L,t)&=&0.
\eea

The probability of polymer survival in the pore is
\be\label{a4}
S(t)=\int_0^L\md s \;c(s,t).
\ee
The translocation probability can be thus expressed as $P_{\rm tr}(t)=1-S(t)$ and the mean-first passage time distribution is therefore $\psi(t)=P'_{\rm tr}(t)=-S'(t)$, or
\be
\label{a5}
\psi(t)=-\int_0^L\md s \;\del_tc(s,t).
\ee
Thus, the translocation time corresponding to the mean-first passage time reads
\be
\label{a6}
\tau\equiv\int_0^\infty\md t\;\psi(t)t=-\int_0^L\md s\int_0^\infty\md t\;\del_tc(s,t)t.
\ee

We define now the transient part of the polymer density function
\be
\label{a7}
u(s,t)=c(s,t)-c_{\rm st}(s),
\ee
with the steady-state polymer probability satisfying the equation $\del_s^2c_{\rm st}(s)+\beta\del_s\left[U'_{\rm p}(s)c_{\rm st}(s)\right]=0$. Thus, the transient solution~(\ref{a7}) equally satisfies the FP equation~(\ref{fp1}),
\be\label{a8}
\del_t u(s,t)=D_{\rm p}\del_s^2 u(s,t)+\beta D_{\rm p}\del_s\left[U_{\rm p}'(s)u(s,t)\right].
\ee
Next, we introduce the Laplace transform of Eq.~(\ref{a7}), 
\be
\label{a9}
Y(s,q)=\int_0^\infty\md te^{-qt}u(s,t).
\ee
After an integration by part, the translocation time~(\ref{a6}) becomes
\be
\label{a10}
\tau=\int_0^L\md s\int_0^\infty\md t\;u(s,t)=\int_0^L\md s\;Y_0(s),
\ee
where we defined $Y_0(s)=Y(s,q=0)$.

According to Eq.~(\ref{a8}), given the initial condition~(\ref{a2}), the Laplace transform $Y_0(s)$ solves the differential equation
\be\label{a11}
D\partial_s^2Y_0(s)+\beta D\partial_s\left[U'_{\rm p}(s)Y_0(s)\right]=-\delta(s)-c_{\rm st}(s).
\ee
Integrating Eq.~(\ref{a11}) around the point $s=0$ and taking into account the vanishing polymer probability for $z<0$ outside the pore, one gets 
\be\label{a12}
Y'_0(0^+)+\beta U'_{\rm p}(0^+)Y_0(0^+)=-\frac{1}{D_{\rm p}}.
\ee
 Accounting now for the absorbing BC~(\ref{a3}), the homogeneous solution to Eq.~(\ref{a11}) follows as
\be
\label{a13}
Y_0(s)=b\;e^{-\beta U_{\rm p}(s)}\frac{\int_s^L\md s'e^{\beta U_{\rm p}(s')}}{\int_0^L\md s'e^{\beta U_{\rm p}(s')}}
\ee
where $b$ is an integration constant. Injecting the solution~(\ref{a13}) into Eq.~(\ref{a12}), one finds $b=\int_0^L\md s'e^{\beta U_{\rm p}(s')}/D_{\rm p}$ and 
\be
\label{a14}
Y_0(s)=\frac{1}{D_{\rm p}}e^{-\beta U_{\rm p}(s)}\int_s^L\md s' e^{\beta U_{\rm p}(s')}.
\ee
Finally, the substitution of Eq.~(\ref{a14}) into Eq.~(\ref{a10}) yields the translocation time~(\ref{12}) of the main text. 

\section{Derivation of the polymer interaction potential $\Delta\Omega_{\rm p}(s)$}
\label{ap3}

In this appendix, we explain the calculation of the polymer-membrane interaction potential $\Delta\Omega_{\rm p}(s)$ in Eq.~(\ref{2}) from the total polymer grand potential 
\be\label{a30}
\Omega_{\rm p}(s)=\Omega_{\rm pm}(s)+\Omega_s(s).
\ee
In Eq.~(\ref{a30}), the term $\Omega_{\rm pm}(s)$ is the interaction energy between the polymer and membrane charges. The second component $\Omega_s(s)$ corresponds to the polymer self-energy accounting for the polarization forces induced by the dielectric contrast between the membrane and the solvent. Below, we review briefly the computation of these two potential components previously derived in Ref.~\cite{Buy2016}. The electrostatic potential $\Delta\Omega_{\rm p}(s)$ will be obtained from the grand potential~(\ref{a30}) at the end of Section~\ref{pols}.

\subsection{Polymer-membrane coupling energy $\Omega_{\rm pm}(s)$}

In Eq.~(\ref{a30}), the grand potential component taking into account the polymer-membrane charge coupling is
\be\label{a32}
\beta\Omega_{\rm pm}(s)=\int\mathrm{d}\br\sigma_{\rm p}(\br)\phi_{\rm m}(\br)
\ee
where we introduced the polymer charge density function
\be
\label{a33}
\sigma_{\rm p}(\br)=-\lc\delta(\br_\pa)\left[\theta(-z)\theta(z+l)+\theta(z-d)\theta(d+s-z)\right].
\ee
The first and second terms inside the bracket of Eq.~(\ref{a33}) correspond to the cis and trans portions of the polymer with the respective lengths $l=L-s$ and $s$. Then, in Eq.~(\ref{a32}), the average electrostatic potential  $\phi_{\rm m}(\br)=\phi_{\rm m}(z)$ induced by the membrane charges satisfies the linear PB equation
\be
\label{a34}
\left[\del_z\e(z)\del_z-\kappa^2(z)\right]\phi_{\rm m}(z)=-4\pi\ell_{\rm B}\sigma_{\rm m}\left[\delta(z)+\delta(z-d)\right].
\ee
In Eq.~(\ref{a34}), the dielectric permittivity and ionic screening functions read
\bea\label{a35}
\e(z)&=&\e_{\rm w}\left[\theta(-z)+\theta(z-d)\right]+\e_{\rm m}\theta(z)\theta(d-z);\\
\label{a36}
\kappa^2(z)&=&\kappa_{\rm b}^2\left[\theta(-z)+\theta(z-d)\right].
\eea
Solving Eq.~(\ref{a34}) with the continuity condition $\phi_{\rm m}(z^-)=\phi_{\rm m}(z^+)$, and the jump condition $\e(z^+)\phi'(z^+)-\e(z^-)\phi'(z^-)=4\pi\ell_{\rm B}\e_{\rm w}\sigma_{\rm m}$ at the charged boundaries located at $z=0$ and $z=d$, one obtains
\be
\label{a37}
\phi_{\rm m}(z)=\frac{2}{\mu\kappa_{\rm b}}\left\{e^{\kappa_{\rm b}z}\theta(-z)+\theta(z)\theta(d-z)+e^{-\kappa_{\rm b}(z-d)}\theta(z-d)\right\}\mbox{sign}(\sigma_m),
\ee
with the Gouy-Chapman length $\mu=1/(2\pi\ell_{\rm B}|\sigma_{\rm m}|)$. Substituting Eqs.~(\ref{a33}) and~(\ref{a37}) into Eq.~(\ref{a32}), the polymer-membrane charge coupling potential finally becomes
\be\label{a38}
\beta\Omega_{\rm pm}(s)=-\frac{2\lc}{\mu\kappa_{\rm b}^2}\left[1-e^{-\kappa_{\rm b}\left(L-s\right)}+1-e^{-\kappa_{\rm b}s}\right]\mbox{sign}(\sigma_m).
\ee

\subsection{Polymer self-energy $\Delta\Omega_s(s)$ and total electrostatic polymer potential $\Delta\Omega_{\rm p}(s)$}
\label{pols}

The polymer self-energy component of Eq.~(\ref{a30}) is given by
\be
\label{a39}
\beta\Omega_s(s)=\frac{1}{2}\int\mathrm{d}\br\mathrm{d}\br'\sigma_{\rm p}(\br)v(\br,\br')\sigma_{\rm p}(\br'),
\ee
where the electrostatic kernel solves the DH equation
\be
\label{a310}
\left[\nabla\cdot\e(z)\nabla-\e(z)\kappa^2(z)\right]v(\br,\br')=-\frac{e^2}{k_{\rm B}T}\delta(\br-\br').
\ee
Exploiting the planar symmetry and Fourier-expanding the kernel as
\be
\label{a311}
v(\br,\br')=\int\frac{\mathrm{d}^2\bk}{4\pi^2}e^{i\bk\cdot(\br_\pa-\br'_\pa)}\tv(z,z'),
\ee
the kernel Eq.~(\ref{a310}) takes the one dimensional form
\be\label{a312}
\left[\del_z\e(z)\del_z-p^2(z)\right]\tv(z,z')=-\frac{e^2}{k_{\rm B}T}\delta(z-z')
\ee
where $p(z)=\sqrt{\kappa^2(z)+k^2}$. The homogeneous solution of the linear differential equation~(\ref{a312}) reads
\bea\label{a313}
\tv(z,z')&=&b_1e^{p_{\rm b}z}\theta(z'-z)+\left[b_2e^{p_{\rm b}z}+b_3e^{-p_{\rm b}z}\right]\theta(z-z')\theta(-z)\\
&&+\left[b_4e^{kz}+b_5e^{-kz}\right]\theta(z)\theta(d-z)+b_6e^{-p_{\rm b}z}\theta(z-d),\nonumber
\eea
for the charge source located at $z'<0$, and 
\bea\label{a314}
\tv(z,z')&=&c_1e^{p_{\rm b}z}\theta(z'-z)+\left[c_2e^{p_{\rm b}z}+c_3e^{-p_{\rm b}z}\right]\theta(z-z')\theta(-z)\\
&&+\left[c_4e^{kz}+c_5e^{-kz}\right]\theta(z)\theta(d-z)+c_6e^{-p_{\rm b}z}\theta(z-d),\nonumber
\eea
for $z'>0$. In Eqs.~(\ref{a313}) and~(\ref{a314}), the coefficients $b_i$ and $c_i$ are integration constants. These constants are to be determined by imposing the continuity of the kernel $\tv(z,z')$ and the displacement field $\e(z)\del_z\tv(z,z')$ at $z=0$ and $z=d$, and by accounting for the additional relations $\tv(z'_-,z')=\tv(z'_+,z')$ and $\del_z\tv(z,z')|_{z=z'^+}-\del_z\tv(z,z')|_{z=z'^-}=-4\pi\ell_{\rm B}$ at the location of the source ion. After some long algebra, the Fourier-transformed kernel takes the form
\be
\label{a315}
\tv(z,z')=\tv_{\rm b}(z-z')+\delta\tv(z,z').
\ee
In Eq.~(\ref{a315}), the first term is the Fourier transformed bulk DH kernel $\tv_{\rm b}(z-z')=(2\pi\ell_{\rm B}/p_{\rm b})e^{-p_{\rm b}|z-z'|}$. The second term corresponds to the  dielectric part of the Green's function originating from the presence of the membrane. This dielectric component reads
\be
\label{a316}
\delta\tv(z,z')=\frac{2\pi\ell_{\rm B}}{p_{\rm b}}\frac{\Delta\left(1-e^{-2kd}\right)}{1-\Delta^2e^{-2kd}}e^{p_{\rm b}(z+z')},
\ee
for $z\leq0$ and $z'\leq0$, 
\be
\label{a317}
\delta\tv(z,z')=\frac{2\pi\ell_{\rm B}}{p_{\rm b}}\frac{\Delta\left(1-e^{-2kd}\right)}{1-\Delta^2e^{-2kd}}e^{p_{\rm b}(2d-z-z')},
\ee
for $z\geq d$ and $z'\geq d$, and
\be
\label{a318}
\delta\tv(z,z')=\frac{2\pi\ell_{\rm B}}{p_{\rm b}}\frac{(1-\Delta^2)e^{(p_{\rm b}-k)d}+\Delta^2e^{-2kd}-1}{1-\Delta^2e^{-2kd}}e^{-p_{\rm b}|z-z'|},
\ee
for $z'\leq0$ and $z\geq d$, or $z'\geq d$ and $z\leq0$. In Eqs.~(\ref{a316})-(\ref{a318}), we defined the dielectric jump function
\be\label{a319}
\Delta=\frac{\e_{\rm w}p_{\rm b}-\e_{\rm m}k}{\e_{\rm w}p_{\rm b}+\e_{\rm m}k}.
\ee

The net interaction potential $\Delta\Omega_{\rm p}(s)$ between the polymer and the charged dielectric membrane corresponds to the grand potential~(\ref{a30}) minus its bulk value.  In the bulk reservoir where there is no charged interface, i.e. $\sigma_{\rm m}=0$, the polymer-membrane charge coupling energy~(\ref{a38}) naturally vanishes. Consequently, the interaction potential becomes
\be
\label{op}
\Delta\Omega_{\rm p}(s)=\Omega_{\rm pm}(s)+\Delta\Omega_s(s),
\ee
where the polymer self-energy renormalized by its bulk value is
\be
\label{a320}
\beta\Delta\Omega_s(s)=\frac{1}{2}\int\mathrm{d}\br\mathrm{d}\br'\sigma_{\rm p}(\br)\left[v(\br,\br')-v_{\rm b}(\br-\br')\right]\sigma_{\rm p}(\br').
\ee
Using Eqs.~(\ref{a315})-(\ref{a318}) in Eq.~(\ref{a320}), after lengthy algebra, the self-energy  Eq.~(\ref{a320}) becomes
\be
\label{a321}
\Delta\Omega_s(s)=\Delta\Omega_{\rm intra}(s)+\Delta\Omega_{\rm inter}(s),
\ee
with the individual self-energy of the polymer portions on the \textit{cis} and \textit{trans} sides of the membrane
\be
\label{a322}
\beta\Delta\Omega_{\rm intra}(s)=\frac{\ell_{\rm B}\lc^2}{2}\int_0^\infty\frac{\mathrm{d}kk}{p_{\rm b}^3}\frac{\Delta\left(1-e^{-2kd}\right)}{1-\Delta^2e^{-2kd}}
\left\{\left[1-e^{-p_{\rm b}s}\right]^2+\left[1-e^{-p_{\rm b}(L-s)}\right]^2\right\},
\ee
and the energy of interaction between the \textit{cis} and \textit{trans} portions of the polymer
\bea
\label{a323}
\beta \Delta\Omega_{\rm inter}(s)=\ell_{\rm B}\lc^2\int_0^\infty\frac{\mathrm{d}kk}{p_{\rm b}^3}\left\{\frac{\left(1-\Delta^2\right)e^{(p_{\rm b}-k)d}}{1-\Delta^2e^{-2kd}}-1\right\}
e^{-p_{\rm b}d}\left[1-e^{-p_{\rm b}s}\right]\left[1-e^{-p_{\rm b}(L-s)}\right].
\eea
The net interaction potential~(\ref{op}) can be finally expressed in terms of the energy components in Eq.~(\ref{a38}), (\ref{a322}), and~(\ref{a323}) as
\be
\label{a324}
\Delta\Omega_{\rm p}(s)=\Omega_{\rm pm}(s)+\Delta\Omega_{\rm intra}(s)+\Delta\Omega_{\rm inter}(s).
\ee

\end{document}